# Mars Astrobiological Cave and Internal habitability Explorer (MACIE): A New Frontiers Mission Concept

By: C. Phillips-Lander (charity.lander@swri.edu)[1], A. Agha-mohamamdi[2], J. J. Wynne[3], T. N. Titus[4], N. Chanover[5], C. Demirel-Floyd[6], K. Uckert[2], K. Williams[4], D. Wyrick[1], J.G. Blank[7,8], P. Boston[8], K. Mitchell[2], A. Kereszturi[9], J. Martin-Torres[10,11], S. Shkolyar[12], N. Bardabelias[13], S. Datta[14], K. Retherford[1], Lydia Sam[11], A. Bhardwaj[11], A. Fairén[15,16], D. Flannery[17], R. Wiens[17]

[1]Southwest Research Institute
[2]NASA Jet Propulsion Laboratory
[3]Northern Arizona University
[4]U.S. Geological Survey
[5]New Mexico State University
[6]University of Oklahoma
[7]Blue Marble Space Institute of Science
[8]NASA Ames Research Center
[9]Konkoly Thege Miklos Astronomical Institute, Budapest, Hungary
[10]Instituto Andaluz de Ciencias de la Tierra (CSIC-UGR), Spain
[11]University of Aberdeen, United Kingdo
[12]USRA/NASA Goddard
[13]University of Arizona
[14]University of Texas-San Antonio
[15]Centro de Astrobiogía, Spain
[16]Cornell University
[17]Queensland University for Technology, Australia
[18]Los Alamos National Laboratory

Cosigners

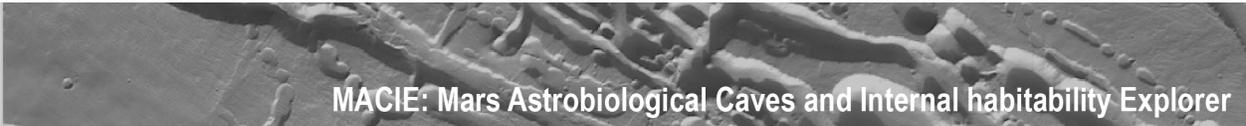
MACIE: Mars Astrobiological Caves and Internal habitability Explorer

**Summary of Key Points**
1. Martian subsurface habitability and astrobiology can be evaluated via a lava tube cave, without drilling.
2. MACIE addresses two key goals of the Decadal Survey (2013-2022) and three MEPAG goals.
3. New advances in robotic architectures, autonomous navigation, target sample selection, and analysis will enable MACIE to explore the Martian subsurface.

### 1. Martian lava tubes are one of the best places to search for evidence of life

The Mars Astrobiological Cave and Internal habitability Explorer (MACIE) mission concept is named for Macie Roberts, one of NASA's 'human computers' (Conway 2007). MACIE would access the Martian subsurface via a lava tube. Lava tube caves are compelling subsurface astrobiological targets because they have their own relatively stable microclimates, are shielded from radiation and harsh surface conditions, and may contain water ice (Blank et al., 2020). They may also provide access to materials vital to future human exploration and in situ resource utilization (ISRU) activities (Sam et al., 2020). **A cave mission represents a compelling next-step in Mars astrobiology and habitability exploration, because it would examine the deeper subsurface (>20x deeper than prior surface craft) without the cost and risk associated with a deep drilling payload.**

#### Subsurface Mars among the best places to look for post-Noachian habitable environments

Post-Noachian Mars (<3.1 Ga) near-surface habitability rapidly degraded, making near-surface environments inhospitable to life as we know it. Organics observed on Mars to date are oxidized due to radiation making it challenging to search for unambiguous biosignatures in near-surface environments (Eigenbrode et al., 2018). Additionally, transient liquid brines may form at the surface, but may not persist long enough to be habitable (Rivera-Valentin et al., 2020). In contrast, modelling of subsurface environments indicates metastable water ice (Williams et al., 2010) and brines (Burt and Knauth, 2003) may be present over extended time periods and have chemical disequilibria necessary to support life. Thus, **subterranean Mars may be the best place to look for habitable environments and evidence of life. NASEM's Astrobiology Strategy (2019) recommends targeting the Martian subsurface habitability, which has yet to be the focus of a planetary mission.**

#### Tharsis lava tubes may have been habitable in the Amazonian

Hundreds of lava tube entrances are located within Hesperian–Amazonian terranes in the Tharsis region (Cushing et al., 2015) and are expected to host metastable water ice (Williams et al., 2010). Mapping of the Tharsis region demonstrates both glacio-volcanism (Cassinelli and Head, 2019) and hydrothermal fluvial activity (Hargitai and Guilick, 2018), which may have occurred into the late Amazonian. This suggests hydrothermal fluid fluxes from the surface to the subsurface may have occurred as they do in modern volcanic systems, as well as melting of glaciers associated with glacio-volcanic activity, similar to Iceland. The Tharsis region also hosts vast fracture networks, which would provide means to transport surface waters and nutrients into the subsurface (Bouley et al., 2018). Thus, **lava tube caves may represent a modern and/or recently habitable environment on Mars. Detection of habitable conditions in the subsurface of Amazonian Mars would fundamentally change our view of Mars' habitability and astrobiological potential through geologic time.**

### 2. MACIE will determine the Habitability and Astrobiological Potential of a lava tube cave

MACIE focuses on two key questions: (1) how long were habitable conditions maintained within a Martian cave? and 2) did life ever colonize a Martian cave? These questions are translated into two primary goals: 1) assess the present and past habitability of a Martian lava tube and (2) search for evidence of past or present life in a Martian lava tube. **MACIE would address multiple goals of**

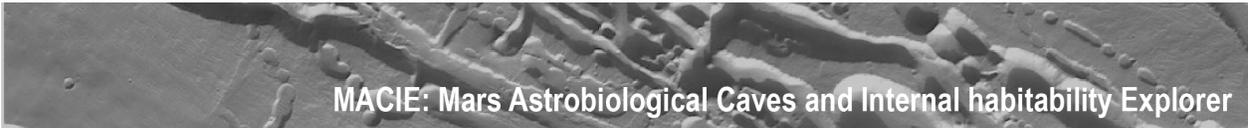

NASA's Strategic Plan (2020), Planetary Decadal Survey (2013-2022), NASEM's Astrobiology Strategy (2019), and MEPAG goals (Bandfield et al., 2020) (**Table 1: Science Traceability Matrix; STM**). MACIE may yield bonus science including insight into future human exploration (MEPAG Goal 4).

### Goal 1: Assess the present and past habitability of a Martian lava tube

Understanding the present and past habitability of a lava tube will be essential in providing the context for evidence of life and/or life-related processes if they are observed. If no evidence of present or past life is detected, the habitability assessment would help explain why evidence of life was not observed. MACIE's habitability assessment addresses the Planetary Decadal Survey Goals 4 and 5, NASEM's Astrobiology Strategy (2019), and multiple MEPAG Goals.

**Objective 1A Determine whether brines or water ice are present:** **Water is essential to habitability of environments as we know them.** MACIE would quantify and characterize the distribution of liquid brines and water ice that may be present in the cave using Raman and visible and infrared reflectance (VISIR) spectroscopy and a camera with appropriate lighting. MACIE would also assess the thickness of the overburden, temperature, and relative humidity along a transect within the cave using a meteorological suite, which can aid in determining why brines or ice are or are not detected during MACIE's mission. Such an endeavor would also help to ascertain the structural integrity and approachability of a particular lava tube or cave for sustained future human exploration. These measurements address MEPAG Goal 2 "Assess the processes and climate of Mars" and Goal 4 "Prepare for human exploration."

**Objective 1B Determine whether aqueous alteration occurred now or in the past:** Evidence of water:rock interaction will determine whether liquid water was present in the cave over geologic time. Using a multispectral instrument, we would determine the presence of alteration minerals, including clays, Fe-oxyhydroxides, sulfates, and other minerals indicative of aqueous alteration of the primary igneous rock substrate. Using Raman and VISIR spectroscopy, MACIE is designed to assess the mineralogy and chemistry of the cave ceilings, walls, and floors and enable MACIE to determine whether oxidant rich dust has been transported from the surface. Martian dust, which contains salts (i.e. chlorides, perchlorate, and sulfates), may influence the sublimation of ice and the development of brines, which would influence the development of redox gradients and long-term maintenance of ice that could support or sustain life. MACIE addresses "the spatial and temporal distribution of potentially habitable environments…in the subsurface," (NASEM, 2019) and MEPAG Goals 2 and 3).

**Objective 1C Determine the presence of nutrients and chemical disequilibria necessary to support life:** The distribution and availability of CHNOPS+Fe elements in the solid phase has been shown to influence microbial colonization in terrestrial lava tubes (Popa et al., 2012; Phillips-Lander et al., 2020). Nutrients required for life are accessible from both the aqueous (condensation, liquid water) and solid phases (rocks, minerals, and water ice). A camer paired with Raman and laser-induced breakdown spectroscopy (LIBS) will identify and quantify the chemistry of ices, primary rock substrates, and alteration phases to determine the availability of nutrients and energy, in the form of chemical disequilibria, available within the cave to promote and sustain life in the absence of light (NASEM, 2019; MEPAG Goal 3).

**Objective 1D Radiation Flux in the Subsurface:** While some microorganisms survive exposure to high radiation levels (Battista, 1997), many microorganisms cannot. Cave roof thickness determines whether the radiation environment is within the cave. We anticipate cosmic rays and neutrons will penetrate a Martian cave with an overburden up to 500 g cm$^{-2}$ (~3 m; Turner and Kunkel, 2017). Quantifying radiation levels using a radiation sensor would serve to explain microbial habitat suitability (MEPAG Goal 1).

Mission Concept for Mars Subsurface Life and Habitability

| | MEPAG GOALS<br>1. Determine if Mars ever supported life;<br>2. Understand the processes and climate of Mars; 3. Understand Mars as a geologic system; and 4. Prepare for human exploration | National Academies Astrobiology Strategy (NASEM, 2019)<br>"What is the spatial and temporal distribution of potentially habitable environments on Mars, especially in the subsurface?" | Decadal Survey (2013-2022) Planetary Habitats Priority Questions<br>"4. Did Mars [...] host ancient aqueous environments conducive to early life, and is there evidence that life emerged?<br>5. Are there contemporary habitats with the necessary conditions, organic matter, water, energy, and nutrients to sustain life, and do organisms live there now?" |
|---|---|---|---|

| Goal | Science Objective | Physical Parameters | Observables with Uncertainty | Instrument | Closure/Benefit |
|---|---|---|---|---|---|
| 1. Assess the present and past habitability of the destination lava tube cave. | 1A. Determine whether habitable conditions have ever existed within the candidate Martian lava tube as quantified by the presence of water. | 1A.1 Determine whether the destination cave atmosphere is consistent with the metastability of water ice (e.g. Williams et al., 2010) | Measure air and wall temperature (+/- 2°C); relative humidity (+/- 5%); atmospheric pressure to (+/- 1 Pa) | Environmental station | Build climate models to understand the factors (climate, environmental, and geologic) influencing the metastability of water ice (MEPAG Goal 2, 3). Bonus: MEPAG Goal 4 |
| | | | Measure radiation levels (0.05 mSV/h) at different depths within the cave to estimate overburden thickness to (+/-1 cm) | Radiation sensor | |
| | | | Map the cave in 3D to build a model of its volume and extent | Camera | |
| | | 1A.2 Determine whether evidence is consistent with the presence of brine or water ice in the destination cave (Hargitai and Guilick, 2018; Williams et al., 2010) | Quantify the spatial distribution of $H_2O$ (brines or solid phase water) in the subsurface on cm-m scale. | Raman, Microimaging camera & Camera | Determine whether the subsurface environment contains brines or water ice (Decadal Theme 2; NASEM 2019; MEPAG Goal 3). Bonus: MEPAG Goal 4 |
| | 1B. Determine whether habitable conditions have ever existed within the candidate Martian lava tube as quantified by evidence of aqueous alteration | 1B.1 Determine whether evidence indicative of past liquid water or water ice exists in the destination cave as evidenced by aqueous alteration | Quantify alteration minerals (+/- 5 wt% on a 10 cm scale) e.g. clays, Fe-oxides, sulfates, etc. indicative of aqueous alteration of the primary geologic substrate | Raman & VISIR spectroscopy | Understand the alteration history (Decadal Theme 2; NASEM 2019; MEPAG Goal 3) |
| | | | Quantify mineral hydration states (+/- 5 wt% on a 10 cm scale) e.g. clays, Fe-oxides, sulfates, perchlorate, etc. | Raman & VISIR spectroscopy | |
| | 1C. Determine whether habitable conditions ever existed within the candidate Martian lava tube evidence of nutrients and chemical disequilibria | 1C.1 Determine whether evidence indicative of nutrients and energy necessary to support life are present in the destination cave. | Map composition of primary substrate of the cave interior to determine mineralogy (+/- 5 wt%) and elemental (+/-0.1 wt%) composition | Mineralogy: Raman, VISIR spectroscopy microimager; Composition: LIBs | Understand nutrients, energy (redox) available to support life (Decadal Theme 2; NASEM 2019; MEPAG Goal 3) |
| | | | Determine the distribution of nutrients (CHNOPS+Fe) (+/-0.1 wt%) | LIBS | |
| | 1D. Determine whether habitable conditions exist within the candidate Martian lava tube as quantified by radiation attenuation | 1D.1 Determine whether the destination cave is significantly deep to attenuate radiation flux to a level compatible with terrestrial extreomphiles | Measure radiation levels (0.05 mSV/h) at different depths within the cave to constrain the atmospheric exchange rate with the surface | Radiation sensor | Understand the impact the radiation environment has on habitability (Decadal Theme 2; NASEM 2019; MEPAG Goal 3). Bonus: MEPAG Goal 4 |
| 2. Search for evidence of present or past life in the desitination lava tube cave | 2A. Determine whether the candidate lava tube contains evidence consistent with the presence and activity of extant or past life | 2A.1 Determine whether the destination cave contains biomolecular compounds, consistent with the presence of carbon-based life. | Determine the presence and identity of organic molecules (i.e. amino acids, lipids, exopolymeric substances, etc) to 1 ppbv of rock, dust, ice and/or water present in the candidate cave | Time-resolved fluorescence spectroscopy | Detect organic biosignatures in the subsurface (Decadal Survey Theme 1; NASEM 2019; MEPAG Goal 1) |
| | | 2A.2 Determine whether the destination cave contains mineral chemistry consistent with bioalteration | Map composition of alteration minerals within the cave, particularly those spatially co-located with organics to determine mineralogy (+/- 5 wt%) and elemental (+/-0.1 wt%) composition | Mineralogy: Raman & VISIR spectroscopy; Composition: LIBs | Detect mineralogical biosignatures in the subsurface (Decadal Survey Theme 1; NASEM 2019; MEPAG Goal 1) |
| | | 2B.1 Determine whether the destination cave contains morphological features consistent with biosignatures of life | Textural evidence including biovermiculations (spatial patterns of organics and minerals), mineral size and shape (10 μm to 2.5 cm) | Larger structures can be resolved by a microimaging camera (e.g > 60 μm) | Detect morphological biosignatures in the subsurface (Decadal Survey Theme 1; NASEM 2019; MEPAG Goal 1) |

**Table 1:** MACIE Science Traceability Matrix

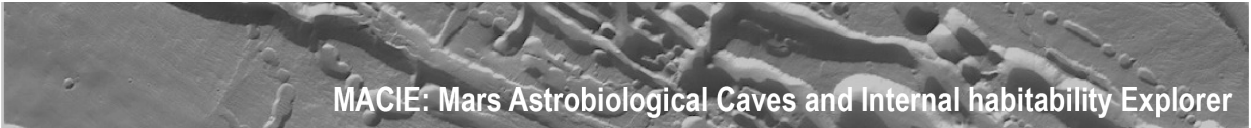

## Goal 2: Search for evidence of extant or past life in a Martian lava tube

MACIE is designed to target multiple objectives to determine whether life may have existed in a lava tube now or in the past, including biomolecule components, minerals formed as byproducts of metabolic activity, and biofabrics (Neveu et al., 2018). These biosignatures would indicate whether life may have colonized a Martian cave and represent a major advance in our understanding of the astrobiological potential of Mars' deep subsurface. These measurements directly address Planetary Decadal Survey Goals 4 and 5 and MEPAG Goal 1 "Determine if Mars ever supported life."

**Objective 2A Detection of biomolecular compounds consistent with carbon-based life:** MACIE is designed to measure the abundances of amino acids, carboxylic acids, lipids, proteins, DNA and RNA, and other organic molecules which may be of biomolecular origin to assess the astrobiological potential of the cave (Neveu et al., 2018) using time-resolved fluorescence spectroscopy.

**Objective 2B Detection of textural evidence indicative of microbial activity:** Microbial mat growth in terrestrial caves can produce biovermiculation patterns, which are micro- to macroscopic deposits resulting from the geochemical interactions of rich microbial colonies with their host rocks' environments (Jones et al., 2008). MACIE is designed to determine whether biofabrics exist in the lava tube using a microimaging camera.

**Objective 2C Presence of alteration products consistent with bioalteration:** Oxide-hydroxide iron and manganese mineral deposits, which are quickly precipitated through microbial redox reactions, can also give rise to distinctive morphologies (Posth et al., 2014). Additionally, Si-rich iron hydroxides, silicate-carbonate assemblages (Mg-rich clays with calcite and aragonite), and amorphous silica layers/films on basalts in terrestrial lava caves are associated with microbial biofilms. These compounds represent the types of chemical and mineralogical biosignatures that may have formed in Martian caves (Léveillé et al., 2007; Miller et al., 2014). Presence of biominerals would be determined with Raman, VISIR, and time resolved fluorescence spectroscopy.

### 3. MACIE Leverages Heritage Instrumentation to Achieve Science Goals

Recent in situ investigations of Mars, including Mars Science Laboratory and Insight, include a suite of instruments to characterize the habitability, composition, and interior structure of Mars.

Access to samples may be limited; many scientifically interesting samples may reside on overhangs or walls. Therefore, **MACIE's instrument suite would conduct stand-off in situ experiments, allowing the mission to pursue the most interesting habitability and astrobiological targets.** The baseline payload includes three instruments: a meteorological suite (e.g. ExoMars 2022 HABIT), a multi-spectral instrument to assess chemistry and mineralogy and provide hand lens quality microscopy (e.g. Mars2020 SuperCam), and a high-resolution context camera (e.g. Mars2020 Mastcam-Z). Estimates for mass, power, and data are based on these heritage instruments listed in **Table 2**. These instruments would address all science experiments in the STM (**Table 1**). **These instruments operate at high stand-off distances from science targets, and represent recently flown, technically mature payloads for a Mars environment that would require limited modifications for a planetary caves mission.**

**Table 2:** Proposed MACIE Instrument Suite

| Instrument | Data (Mb/sol) | Mass (kg) | Power (W) | Heritage |
|---|---|---|---|---|
| Meteorological Suite | 11 | 0.92 | 17 | HABIT |
| Multi-spectral | 4.2 | 5.6 | 17.9 | SuperCam |
| Camera | 148 | 4 | 11.8 | Mastcam-Z |

**Meteorological suite**

HAbitability: Brines, Irradiation and Temperature (HABIT) (Martín-Torres et al., 2020) measures temperature, relative humidity, barometric pressure, wind, dust, and radiation; these data would be incorporated into 2D and 3D cave climate models, explain the presence or absence of predicted water ice, and determine the radiation environment (Science Objectives 1A and 1C; **Table 1**).

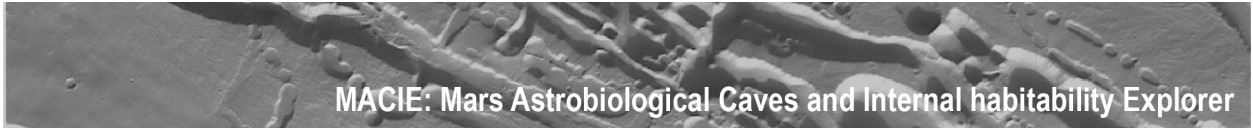

### Multi-spectral Instrument

A multi-spectral instrument similar to the Mars2020 SuperCam suite (with stand-off LIBS, Raman, VISIR, and fluorescence spectroscopies, and color context imaging) would provide multiple probes to assess the habitability and astrobiological potential of interesting targets (Rees et al., 2019). Primary and alteration mineralogy could be characterized using Raman spectroscopy and VISIR. LIBS provides elemental abundances of target material, with potential to yield direct measure of CHNOPS+Fe elements (Rees et al., 2019). These analyses would provide data required to address Science Objective 1B. SuperCam's remote microimaging (RMI) camera has a resolution of 60 μm at 1.5 m stand-off distance, which would allow the detection of ice (Objective 1A) and biofabrics (Objective 2C; **Table 1**). The camera would require active illumination; we baseline a tungsten halogen light with an optical system that projects a flat illumination of 10° for both RMI and the camera (below). Organics could be determined with time-resolved fluorescence spectroscopy as organics fluoresce when excited by the 532 nm laser (Objective 2A; **Table 1**). Fluorescence from organics decays over very short timeframes (<1 ns to 200 ns), significantly shorter than fluorescence from minerals (μs-ms), allowing detection and differentiation of organic and mineral components in the sample.

### Camera

Context imaging would be conducted via a camera like Mastcam-Z, which would allow for high definition and 3D cave images when paired with an illumination source. Image resolutions vary between 150 μm and 750 mm per pixel depending on distance. The camera would be able to capture a range of images to map the extent of brines or ice in the cave and provide high resolution imaging of the cave itself, which would determine cave size and extent (Objective 1A). The 3D information extracted could further be used to characterize the geomorphometric parameters such as orientation, slope, and surface roughness to enable safe rover maneuvering.

### Technological Development

Key technological developments for instrumentation would further aid Mars cave exploration including autonomous sample selection and data processing. Communication between the robotic platform and Earth may be limited due to bandwidth restrictions between the rover and relay points (e.g. deployable relays or orbiting assets); therefore, mission operations may require autonomous sample selection to vet targets before measurement with resource-intensive instruments. Additionally, science data collected throughout the mission may need to be processed or interpreted onboard to prioritize downlinking the most valuable science data in cases of bandwidth-restricted mission architectures. MACIE would benefit from advances in autonomous sample selection using machine-learning algorithms currently being advanced for the Europa Lander and through DoD projects.

## MACIE Mission Architecture and Concept of Operations (ConOps)

### Site Selection

**Cave site selection would be based on expected habitability and astrobiological potential, as defined by smaller entrance opening, longer subsurface cavity lateral extent (Howarth, 1980), and persistence of water ice (Williams et al., 2010).**

Determining subsurface lava tube expression (entrance angle, size, shape, and lateral extent) remains challenging. Roughly 60% of the Mars Global Cave Candidate Catalog remains unimaged by high-resolution instruments. The low roll angle of Mars Reconnaissance Orbiter (MRO ≤ 30°), which hinders the ability of High-Resolution Imaging Science Experiment (HiRISE) to image as obliquely as Lunar Reconnaissance Orbiter Camera (LROC ≤ 40°), combined with the current lack of coverage

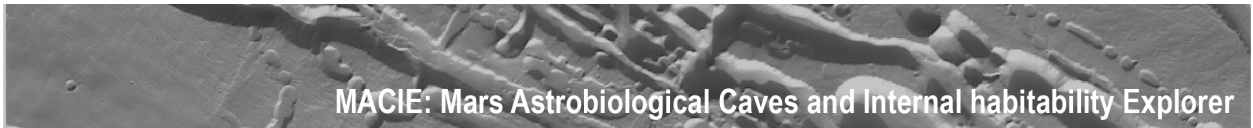

over candidate cave sites may have led to a bias in the available dataset toward vertical entrances. Some horizontal entrances have been identified, which would allow lateral entrance to a cave (Cushing written comm., 2020)

Geophysical methods such as radar, gravity, and magnetic field analyses could help link candidate cave entrances with subsurface void spaces of lateral extent. These methods are sensitive to changes in subsurface structure. Shallow Radar (SHARAD) and Mars Advanced Radar for Subsurface and Ionosphere Sounding (MARSIS) orbital sounders are long-lived experiments with a high combined surface coverage. Preliminary SHARAD results suggest this sounder can detect subsurface

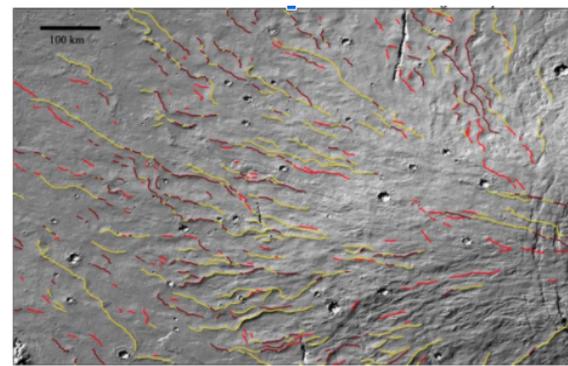

**Figure 1:** Mapped candidate cave entrances and possible lava tube extents at Alba Mons (40.5°N 250.4°E), just north of the Tharsis region indicate lava tubes had shallow slopes <1° (from Perry et al., 2019). This suggests lava tubes will have relatively minimal slopes, enhancing

interfaces at tube ceilings 5-7 m below the surface and with diameters of 2-12 m (Perry et al., 2019). Context camera (CTX) visual and THermal Emission Imaging System (THEMIS) infrared imaging at Alba Mons, just north of the Tharsis region, mapped candidate cave entrances and possible lava tube extents and indicated volcanogenic flanks where lava tubes were observed and had shallow slopes <1°, which suggests cave traverse difficulty would be insignificant, not accounting for the rockiness of the floor material (Perry et al., 2019; **Figure 1**). **Ongoing radar reconnaissance is essential in characterizing lava tube morphologies and extents and would inform MACIE's site selection and design constraints.**

## MACIE Mission Architecture

MACIE's configuration builds on knowledge of subsurface void spaces on Earth, the Moon, and Mars. Several existing mission architecture options could be adapted to target the Martian subsurface (**Figure 2**). Axel, featured in the MoonDiver Discovery Class Mission proposal targeting lunar pits, is a mature technology (technology readiness level; TRL 6) developed to explore vertical entrances known to exist on the Moon and Mars (Nesnas et al., 2020). The ambulatory Spot rover (TRL 5) deployed in a current DARPA SubT competition (Bouman et al., 2020) could be tethered for a vertical entrance. Lateral entrances would not necessitate tethering, reducing risk. Precision landing (>150 m from an entrance) could be used to minimize surface travel prior to exploring the cave.

Remotely operated aerial systems (UAS), like the Mars2020 helicopter, could be used in either a precursor scouting mission (Bapst et al., 2020) or as part of a mission where it is paired with a rover. UAS would survey the area and assess the best cave to explore from risk and science perspectives (Fan et al., 2019; Kanellakis et al., 2020; Sasaki et al., 2020;). In addition, such UAS could be equipped with (1) thermal cameras for detection and comparative analysis of caves and (2) hyperspectral cameras for providing geological/mineralogical information of the surface. UAS, which provide high-resolution oblique images, can also enable 3D terrain modelling with high precision, accuracy, and resolutions (Sam et al., 2020) to enable safe and efficient rover approach to the entrance.

## Robotic and Autonomy Technology Development

Robotic and artificial intelligence technologies targeting mapping, target selection, and chemical analysis have matured sufficiently to prepare us to explore a Martian cave. The DoD's DARPA Subterranean (SubT) Challenge has fueled autonomous technology demonstration for cave exploration (Agha-mohammadi et al., 2019) including (1) mobility in unknown, rugged terrains, narrow passages, and vertical shafts (Bouman et al., 2020), (2) perception (GPS-denied navigation and autonomous motion for >100 m) (Ebadi et al., 2020, Santamaria-Navarro et al., 2019), (3) autonomy (adapting to terrain and assessing risk without ground communication) (Kim et al., 2019; Otsu et al., 2020), and (4) communication for exploring subsurface voids. Additional technological developments in these areas would enhance MACIE's operations and reduce mission cost and risk.

### 5. Cost Justification

Leveraging the precision-landing heritage from Mars2020 and Mars Sample Return, we anticipate **judicious site selection, continuing advancements in robotics, and autonomous sampling and robotic operations would allow MACIE to fit New Frontiers class mission within this decade.** However, on the basis of a preliminary concept study we conducted this year, a true-life detection mission to a Martian cave would exceed a New Frontiers cost cap and bump up to Flagship Mission cost.

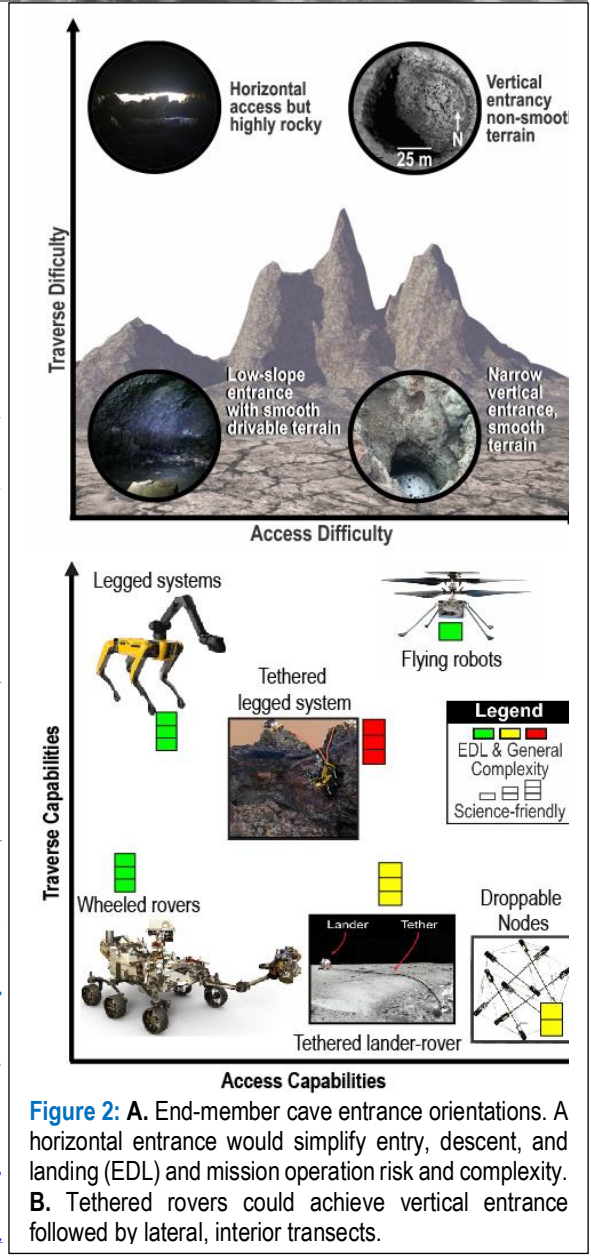

**Figure 2: A.** End-member cave entrance orientations. A horizontal entrance would simplify entry, descent, and landing (EDL) and mission operation risk and complexity. **B.** Tethered rovers could achieve vertical entrance followed by lateral, interior transects.